\documentclass[fleqn,10pt]{wlscirep}
\pdfoutput=1
\usepackage{cite}
\usepackage{graphics}
\usepackage{amsmath}
\usepackage{amssymb}
\usepackage{graphicx}
\usepackage{dcolumn}
\usepackage{epstopdf}
\usepackage{hyperref}

\title{Super-Resolution $^{1}$H Magnetic Resonance Spectroscopic Imaging utilizing Deep Learning}

\author[1]{Zohaib Iqbal}
\author[1]{Dan Nguyen}
\author[2]{Gilbert Hangel}
\author[2]{Stanislav Motyka}
\author[2]{Wolfgang Bogner}
\author[1,*]{Steve Jiang}
\affil[1]{Medical Artificial Intelligence and Automation Laboratory, Department of Radiation Oncology, University of Texas Southwestern Medical Center, Dallas, TX, USA 75390 }
\affil[2]{High Field MR Center, Department of Biomedical Imaging and Image-guided Therapy, Medical University of Vienna, Christian Doppler Laboratory for Clinical Molecular MR Imaging, Austria}
\affil[*]{Corresponding author contact: steve.jiang@utsouthwestern.edu }

\begin{abstract}
Magnetic resonance spectroscopic imaging (SI) is a unique imaging technique that provides biochemical information from \textit{in vivo} tissues. The $^{1}$H spectra acquired from several spatial regions are quantified to yield metabolite concentrations reflective of tissue metabolism. However, since these metabolites are found in tissues at very low concentrations, SI is often acquired with limited spatial resolution. In this work we test the hypothesis that deep learning is able to upscale low resolution SI, together with the T1-weighted (T1w) image, to reconstruct high resolution SI. We report on a novel densely connected Unet (D-Unet) architecture capable of producing super-resolution spectroscopic images. The inputs for the D-UNet are the T1w image and the low resolution SI image while the output is the high resolution SI. The results of the D-UNet are compared both qualitatively and quantitatively to simulated and \textit{in vivo} high resolution SI. It is found that this deep learning approach can produce high quality spectroscopic images and reconstruct entire $^{1}$H spectra from low resolution acquisitions, which can greatly advance the current SI workflow.  
\end{abstract}
\begin{document}

\flushbottom
\maketitle
%
%
\thispagestyle{empty}

\noindent 
\section*{Introduction}

\setlength{\parskip}{0pt}
Magnetic resonance imaging (MRI) continues to be a versatile modality capable of providing anatomical, metabolic, and functional information from various regions of the body \textit{in vivo}. In particular, magnetic resonance spectroscopic imaging (SI)\cite{brown1982nmr} is able to yield important data regarding the metabolism of different tissues, and has been especially useful for studying the metabolism of the human brain\cite{soares2009magnetic}. Some important biochemicals, or metabolites, in the brain include N-acetyl aspartate (NAA), glutamate (Glu), glutamine (Gln), creatine (Cr), choline (Ch), and myo-Inositol (mI)\cite{govindaraju2000proton}. Each metabolite plays an important role in regulating energy consumption in the brain, and some metabolites also play critical functional roles, including roles as neurotransmitters\cite{ramadan2013glutamate}. It is well known that metabolic changes occur in parallel with anatomical changes for a myriad of pathologies\cite{soares2009magnetic}, and these metabolic changes may even occur before structural changes are detected. While SI has continued to be an active area of research over the past several decades, there are still major roadblocks into standardizing this technique and including it into clinical protocols.  

One of the major disadvantages of SI is the long acquisition duration associated with obtaining spectra from several voxels of interest. This is primarily due to the fact that many of the important metabolites are found in the brain at low concentrations; these metabolites are typically present in the brain at 1-12 mM concentrations\cite{govindaraju2000proton}. Therefore, in order to accurately detect these biochemicals, several signal averages have to be obtained or larger voxel volumes have to be acquired to improve the signal to noise ratio (SNR) for the experiment. As a result, spatial resolution tends to be coarse for many SI sequences. This low resolution, coupled with other technical problems such as partial volume effects, hinders the overall diagnostic capabilities of the SI technique.  

There have been many advances in the technological implementation of SI that allow for faster acquisition and better spatial resolution. One of the primary acceleration methods is echo planar spectroscopic imaging (EPSI)\cite{mansfield1984spatial,posse1995high}, which collects spectral data from an entire line of k-space in a single repetition time (TR) utilizing an echo planar readout. This spatio-spectral acquisition approach has also been applied in non-cartesian SI methods, such as spiral acquisitions\cite{adalsteinsson1998volumetric}, concentric circular acquisitions\cite{furuyama2012spectroscopic}, and rosette acquisitions\cite{schirda2009rosette}. In addition, parallel imaging\cite{pruessmann1999sense,dydak2001sensitivity,griswold2002generalized} can also be used to accelerate the collection of SI data. Sensitivity encoding (SENSE) has been applied in combination with EPSI\cite{otazo2007accelerated} to facilitate even faster acquisition times. Recently, research has also focused on the application of various sampling schemes that allow for reduced scan time\cite{strasser20172+,wilson2015accelerated,iqbal20163d,posse2013mr,lam2014subspace}. Some studies\cite{hingerl2017density,hangel2016ultra} have even demonstrated protocols capable of obtaining spectroscopic images at 64x64 or 128x128 resolution in less than 20 minutes. Although these advances have improved the field significantly, SI is still understandably seen as a low SNR, low resolution technique. 

In order to combat the limits of the experimentally acquired resolution, many post-processing methods have been developed for super-resolution SI\cite{jain2017patch,haldar2008anatomically,hu1988slim,cengiz2017super,jacob2007improved,liang1991generalized,kasten2016magnetic}. These methods have mainly focused on model-based reconstruction methods and regularized reconstruction approaches. While many super-resolution methods are independent of the acquisition protocols, there are some techniques, such as the spectroscopic imaging by exploiting spatio-spectral correlation (SPICE) method\cite{lam2014subspace}, that show reconstruction benefits by employing inter-dependent sequences. Unfortunately, the majority of super-resolution methods either tend to be very complicated to implement, or generally show poor reconstruction results. Since experimental acquisitions have many technical challenges, there is also a large concern over the true gold standard for these super-resolution techniques. Without a true standard of comparison, which is a large problem in the spectroscopic imaging field, many studies qualitatively and quantitatively compare their methods with less ideal standards such as bicubic interpolation. 

Deep learning is an advancing field that has shown extraordinary results for image processing\cite{lecun1989backpropagation,lecun2015deep,krizhevsky2012imagenet}. Convolutional layers and networks are capable of extracting valuable features from images, and can further process these features into labels or other images for classification, segmentation, and other uses. One network that has been extremely beneficial for the field of automated medical imaging segmentation is the UNet\cite{ronneberger2015u}, which allows for a pixel-wise transformation of an input image into an output image. Essentially, deep learning excels at computing an unknown transformation by using a large example dataset, often referred to as a training set. We hypothesize that UNet, or some other deep neural networks are able to upscale low resolution SI (LRSI), together with the T1-weighted (T1w) image, to produce high resolution SI (HRSI). To test this hypothesis the biggest challenge is that a large, publicly available SI dataset is unavailable and difficult to acquire experimentally. In order to create this data set, HRSI (128x128 pixels) and LRSI (16x16 pixels or some other low resolution) experiments would have to be performed on thousands of diverse patients with different pathologies, which is not feasible. Thus, it is seemingly impossible to perform deep learning for super-resolution SI. 

In this paper we report a novel work on the development of a deep learning technology capable of producing super-resolution spectroscopic images. An SI generator is used to produce LRSI and HRSI data in order to train and test a deep learning model. Using this data, a UNet taking advantage of densely connected layers (D-UNet) is built and trained. The inputs for the D-UNet are the T1w image and the low resolution SI image while the output is the high resolution SI. The results of the D-UNet reconstruction are compared both qualitatively and quantitatively to simulated and \textit{in vivo} high resolution SI data.  

\section*{Methods}

\subsection*{Spectroscopic Imaging Dataset}
Two different MRI data sets were utilized to produce synthetic SI data for developing the deep learning model. The first MRI data set comprised of 27 axial slices from the MATLAB MRI dataset. MR images as well as white matter (WM) and gray matter (GM) masks from the open access series of imaging studies (OASIS) project\cite{marcus2007open}, which contained 416 axial images from subjects ranging in age from 18-96 years old, were also used. From these limited data, 102,169 SI datasets were synthesized using an SI generator. Each synthetic dataset consisted of three images: an augmented T1w ($aT1w$) image, an LRSI and an HRSI. The SI generator first randomly allocated a chemical distribution in the WM and GM of the T1w image. This process replicated the fact that a certain chemical may be more present in a particular tissue type. This random distribution lead to the production of an initial matching HRSI. Then, the SI generator used a variety of random transformations which acted on both the T1w image and HRSI: image rotation, image contrast alteration, image field of view (FOV) reduction, and lesion (appearing either hyper-intense or hypo-intense on a T1w image) addition. A further transformation was applied to the HRSI to mimic the fact that a chemical may be more concentrated in the anterior, posterior, left, or right regions of the brain. These processes produced the $aT1w$ image and final HRSI. Finally, the LRSI was produced by down-sampling the HRSI image in the Fourier domain via k-space truncation. It is important to note that since many of the transformations were random, the same input T1w axial image could have several unique outputs, which is why a large number of datasets were able to be produced.

\begin{figure}
	\center
	\hspace*{-0.5cm}
	\includegraphics[scale=0.4]{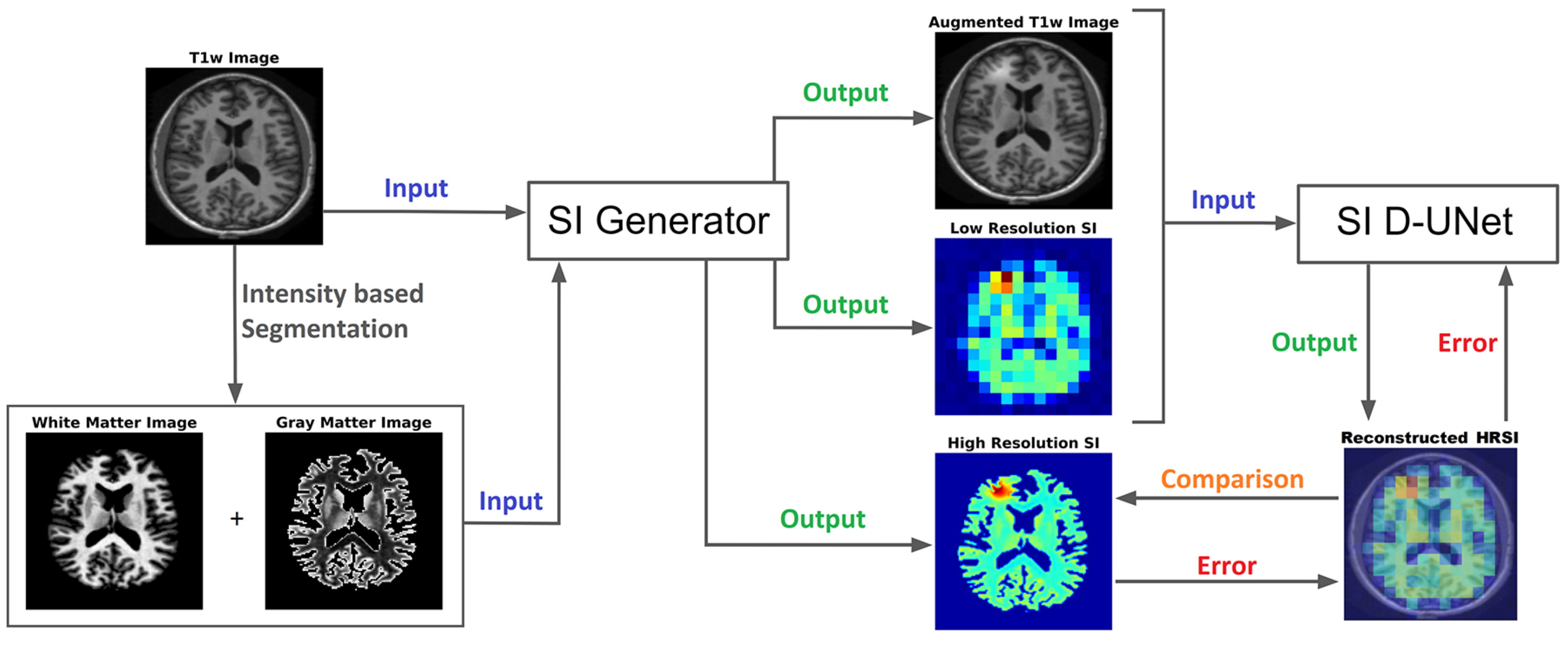}
	\caption{\small The workflow for training the D-UNet model is shown. The SI generator provides a dataset consisting of an augmented T1w image, a low resolution spectroscopic image, and a ground truth high resolution spectroscopic image. Then, the network transforms the $aT1w$ and LRSI into an initial HRSI reconstruction. This reconstruction is compared to the ground truth, and the mean squared error is calculated. Utilizing this error, the model changes the weighting parameters for the features, and continues training by using a different dataset. After training on 102,000 datasets, the model weights are refined and the reconstruction errors are minimized.}
	\label{fig:flow}
\end{figure}

\begin{figure}
	\center
	\includegraphics[scale=0.25]{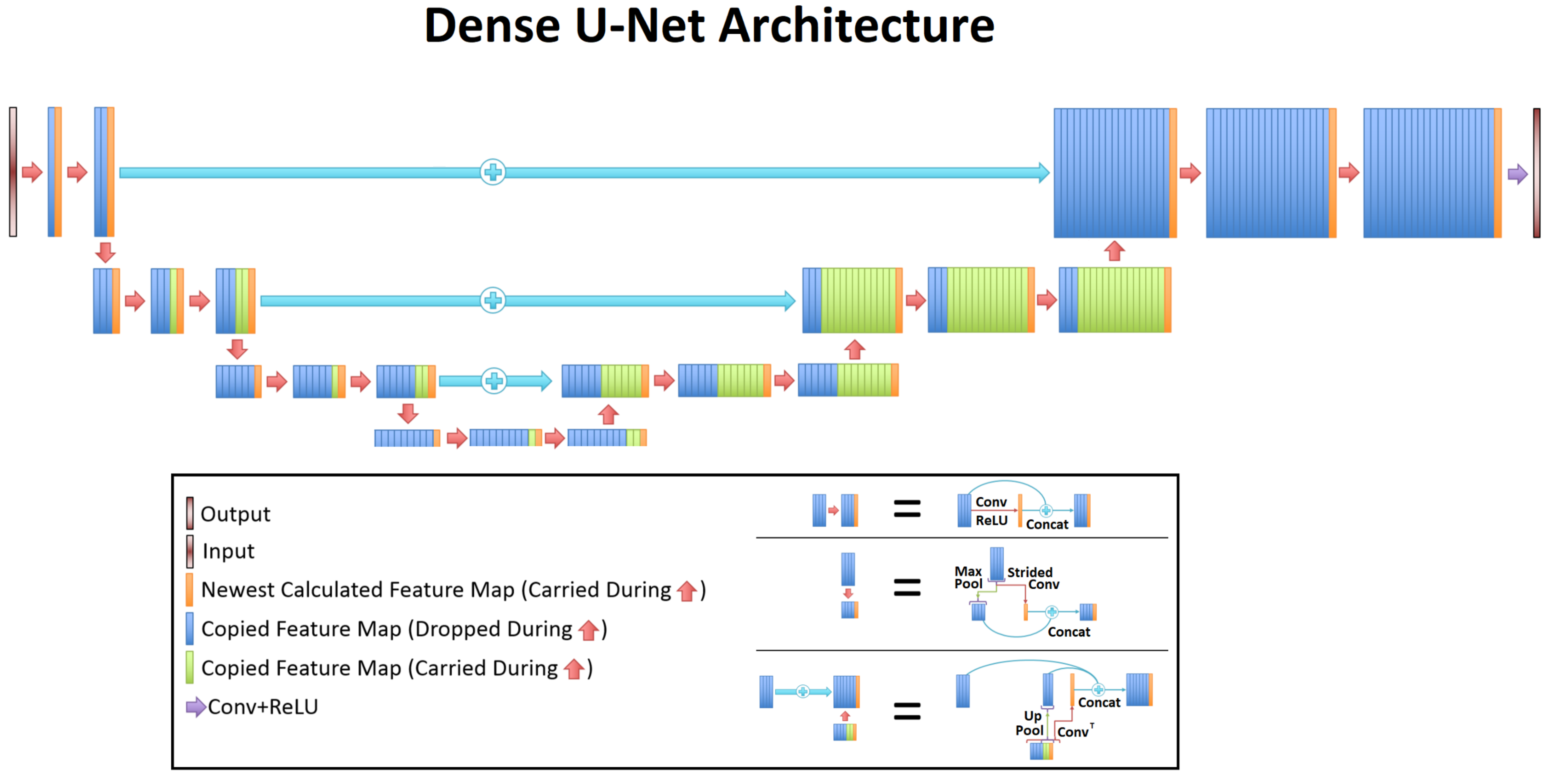}
	\caption{\small The general D-UNet architecture is displayed. Each forward convolution consisted of a convolutional layer and a concatenation process. This concatenation carries over important features which can be used to make the next layer more intelligent. In addition to local concatenations, certain features were concatenated to deeper layers in the network. This allows for prior information to improve the overall reconstruction quality. In order to use the most information possible, the last convolutional layer contains all of the carried over features.}
	\label{fig:dunet}
\end{figure} 

\subsection*{Densely Connected UNet (D-UNet) Architecture and Training }
The UNet architecture\cite{ronneberger2015u} is typically implemented for segmentation purposes, however it primarily operates by performing pixel-wise transformations on input images, which is applicable to the SI super-resolution problem. Using standard convolutional and max pooling layers, the UNet first continuously convolves and pools the input image until the image reaches a small size, which aids in extracting valuable global features. Next the image is scaled up through a combination of up-pooling, transpose convolutions, and feature concatenations. This second process helps to identify vital local features so that the UNet can refine the image at a finer resolution. However, due to the number of features necessary for this process, the classical UNet suffers from extremely long training times, overfitting issues, and potential inefficiencies when tuning the weights. Therefore, this study utilized densely connected convolutional layers\cite{huang2017densely} to develop the novel densely connected UNet (D-UNet) architecture, and the workflow for training is shown in Fig. \ref{fig:flow}. Densely connected networks carry over features from layer to layer, allowing for all previous information to be used for determining important features. The general architecture of the D-UNet used in this study is shown in Fig. \ref{fig:dunet}. The D-UNet utilized 32 feature maps at every max pooling layer. In addition, all convolutional layers made use of the ReLU activation function\cite{krizhevsky2012imagenet} and used a dropout\cite{srivastava2014dropout} of 0.1. Certain features, shown in green and orange in Fig. \ref{fig:dunet}, were copied over to the following layers, and were also concatenated later on in the network. In total, three max pooling layers were used for the D-UNet. Since low resolution SI experiments can have diverse resolutions, three identical D-UNets were made to upscale low resolution spectroscopic images for acquisitions with 16x16, 24x24, and 32x32 spatial points. 

The D-UNet required two inputs: a rescaled (128x128 points) T1w image and the corresponding LRSI image (16x16, 24x24, or 32x32 points) upscaled using nearest-neighbor interpolation (128x128 points). The predicted output of the D-UNet was a denoised HRSI image (128x128 points). For training, $aT1w$, HRSI, and LRSI were created from the SI generator, as described above. The Adam optimizer\cite{kingma2014adam} was used with a learning rate set to 1x10$^{-3}$, and mean squared error (MSE) was used as the cost function, which determined the difference between the D-UNet output and the desired output: 

\begin{equation} \label{eq:mse}
MSE = \sum\sum\frac{(O - SI_{HR})^2}{m^2}
\end{equation}

Above, $O$ is the output of the D-UNet, $SI_{HR}$ is the true simulated high resolution SI, and $m$ is the output dimension of the network, which in this case is 128. The summations are performed over both dimensions to yield a single value. The network was trained on an 8GB Quadro K5200 graphical processing unit (GPU) using the Keras\cite{chollet2015keras} and Tensorflow\cite{abadi2016tensorflow} packages in Python 3.6.

Two datasets were made for the development and evaluation of the three D-UNets: a training dataset and a testing dataset. The training dataset comprised of 102,000 data from the SI generator using 135 axial images. The testing dataset used 169 different axial images (independent from the training set) from the OASIS project, and 169 matched $aT1w$, HRSI, and LRSI images were produced via the SI generator. Each of the three D-UNets were trained for a total of 102 epochs. For this study, an epoch was defined as 1,000 image sets. The first two epochs were trained using a batch size of one to ensure that the network would not fall into a local minimum. The remaining 100 epochs were trained with a batch size of 10.

\subsection*{D-UNet Evaluation and Comparison Metrics }
\subsubsection*{Testing Set Evaluation}
The three D-UNets evaluated all 169 matched images ($aT1w$ and LRSI) to produce reconstructed high resolution spectroscopic images ($Recon_{16x16}$, $Recon_{24x24}$, and $Recon_{32x32}$). These reconstructed images were compared to the ground truth HRSI using mean squared error. This process was repeated with varying noise levels inserted into the input LRSI in order to determine the role of noise on the reconstruction process. These reconstructed images were also compared to zero-filling and bicubic interpolation to assess the improvement of the D-UNet results over standard methods. For this comparison, both zero-filling and bicubic interpolation were applied to an LRSI of 32x32 points to generate the 128x128 interpolated images. 

\subsubsection*{Spectral Reconstruction Evaluation}
In addition, the three D-UNets were used to reconstruct magnitude spectra point-by-point from low spatial resolution to high spatial resolution. Magnitude spectra were used because the model was not trained for evaluating real and imaginary numbers simultaneously. From the test set, a single subject was used to generate high resolution chemical maps of the major metabolites, including NAA, Glu, Gln, Cr, Ch, and mI. GAMMA simulation\cite{smith1994computer} was used to simulate the spectra for these metabolites using an echo time (TE)=30ms, spectral bandwidth of 2000Hz, and time points = 512 for a magnetic field strength ($B_0$) of 3T. Also, the spectra were exponentially line broadened to roughly 8Hz. These spectra were then distributed spatially based on their respective high resolution maps, and were transformed to produce LRSI. The T1w image and LRSI were input into the three D-UNets to produce $Recon_{16x16}$, $Recon_{24x24}$, and $Recon_{32x32}$ spectral data. Two example spectra were extracted from these reconstructed images and compared to the simulated ground truth using mean squared error.  

\subsubsection*{In Vivo Evaluation}
Finally, high resolution spectroscopic images were acquired on a 7T whole-body MR scanner
(Magnetom, Siemens Healthcare, Erlangen, Germany) using a previously published protocol\cite{hangel2016ultra}. The Institutional Review Board (IRB) at the Medical University of Vienna approved the study and ten healthy volunteers (mean age = 31.7 years old) signed written and informed consent forms. All experiments were performed in accordance with relevant guidelines and regulations. The protocol utilized free induction decay based MR spectroscopic imaging (FID-MRSI)\cite{bogner2012high} with TR=200ms for a total scan time of 21 minutes. After acquisition, residual lipids were removed using $\ell_2$ regularization\cite{bilgic2014fast} and the spectra were quantified using the LCModel\cite{provencher1993estimation} package to yield concentrations for several metabolites. Therefore, high resolution (128x128 pixels, 1.7x1.7mm$^2$) metabolite maps for NAA, Cr, Ch, Glu, Gln, and mI were obtained. These metabolite maps were down-sampled to 32x32 resolution images and were input into the 32x32 D-UNet along with corresponding T1w images to yield $Recon_{32x32}$ for all datasets. These reconstructed images were then compared to the experimentally acquired HRSI using mean squared error as described in equation \ref{eq:mse}. In addition, Glu/Cr and Ch/Cr ratios for both the reconstructed and experimentally acquired images were measured over all ten subjects. These ratios were investigated as a function of T1w intensity, which directly corresponds to the ratio of WM and GM in the brain. Finally, correlations between the reconstructed and experimental results were performed to yield the correlation coefficients ($r$) for the Glu/Cr and Ch/Cr ratios.

\subsection*{Data Availability}
The datasets used for this study are freely available from the OASIS project\cite{marcus2007open}. All other data and protocols are included in this article or in a previous publication\cite{hangel2016ultra}.

\section*{Results}

\begin{figure}
	\center
	\hspace*{-1.0cm}
	\includegraphics[scale=0.5]{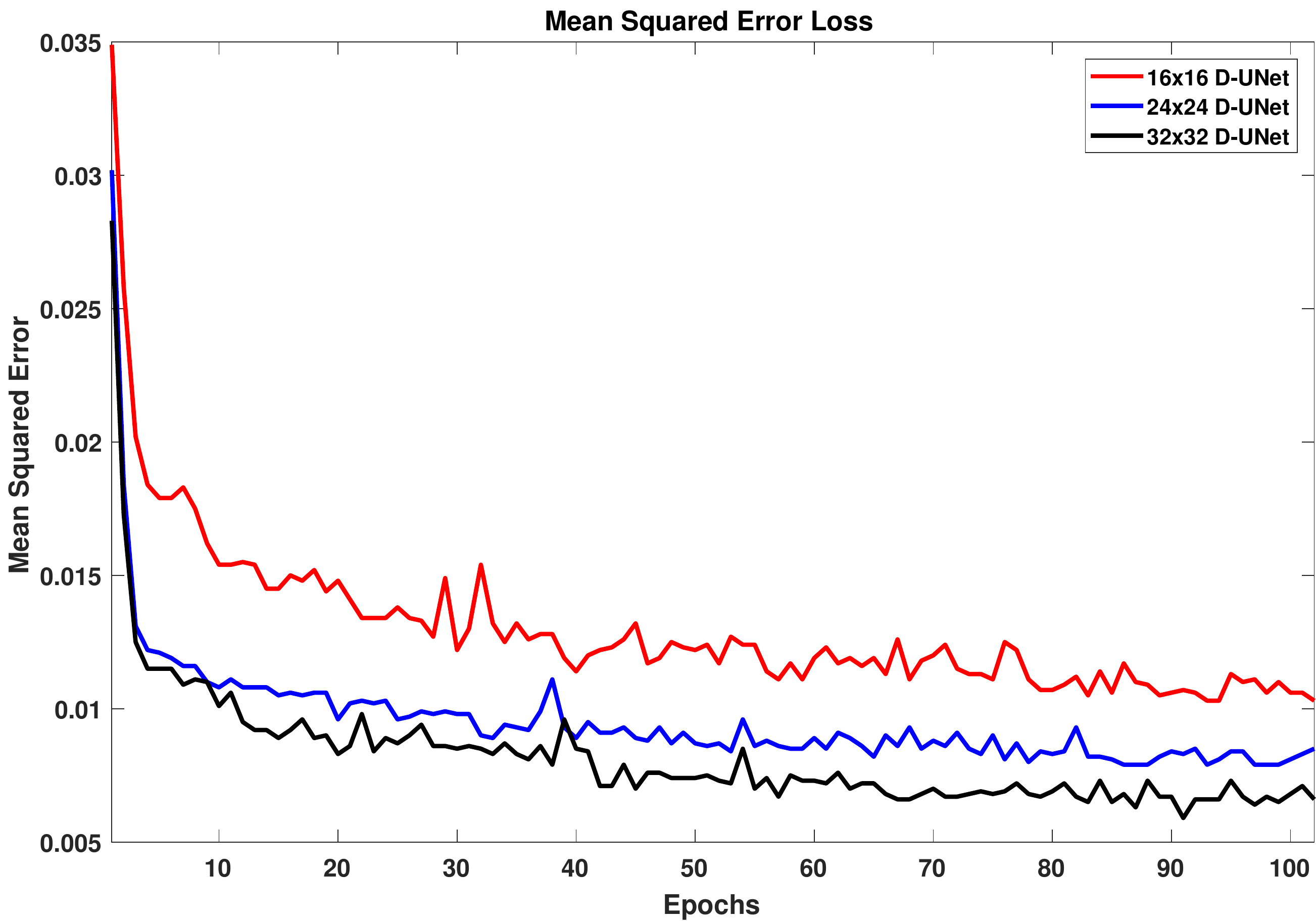}
	
	\caption{\small The loss functions for the 16x16 D-UNet (red), 24x24 D-UNet (blue), and the 32x32 D-UNet (black) are shown. All three loss functions drop significantly in the first 2 epochs, and then gradually decrease as the training continues. Overfitting is not an issue with the current training method, since all of the data are only seen by the network once. While more epochs could be used, the loss function flattens after 70 epochs, which implies that further training will yield minimal improvement.}
	\label{fig:loss}
\end{figure}

\subsection*{Training Results}
Due to the novel D-UNet architecture, the mean squared error loss rapidly converged close to a reasonable value after only 2 epochs for all three networks, and the loss functions are shown in Fig. \ref{fig:loss}. The loss continued to decrease with more epochs when a larger batch size was used for the remaining 100 epochs. From Fig. \ref{fig:loss}, it is clear that the final loss was better for the 32x32 D-UNet than the 24x24 or 16x16 D-UNets. This is theoretically expected because higher initial resolution should aid in the estimation of unknown points, and this is true for conventional resolution enhancement techniques as well. While a low dropout was used in the architecture, overfitting was not a primary concern for the D-UNet training framework because of the reduced number of weighting parameters in the model. The results from the testing dataset also highlight the fact that the D-UNet training was generalized and applicable to never before seen data.   

\begin{figure}
	\center
	\hspace*{-0.5cm}
	\includegraphics[scale=0.3]{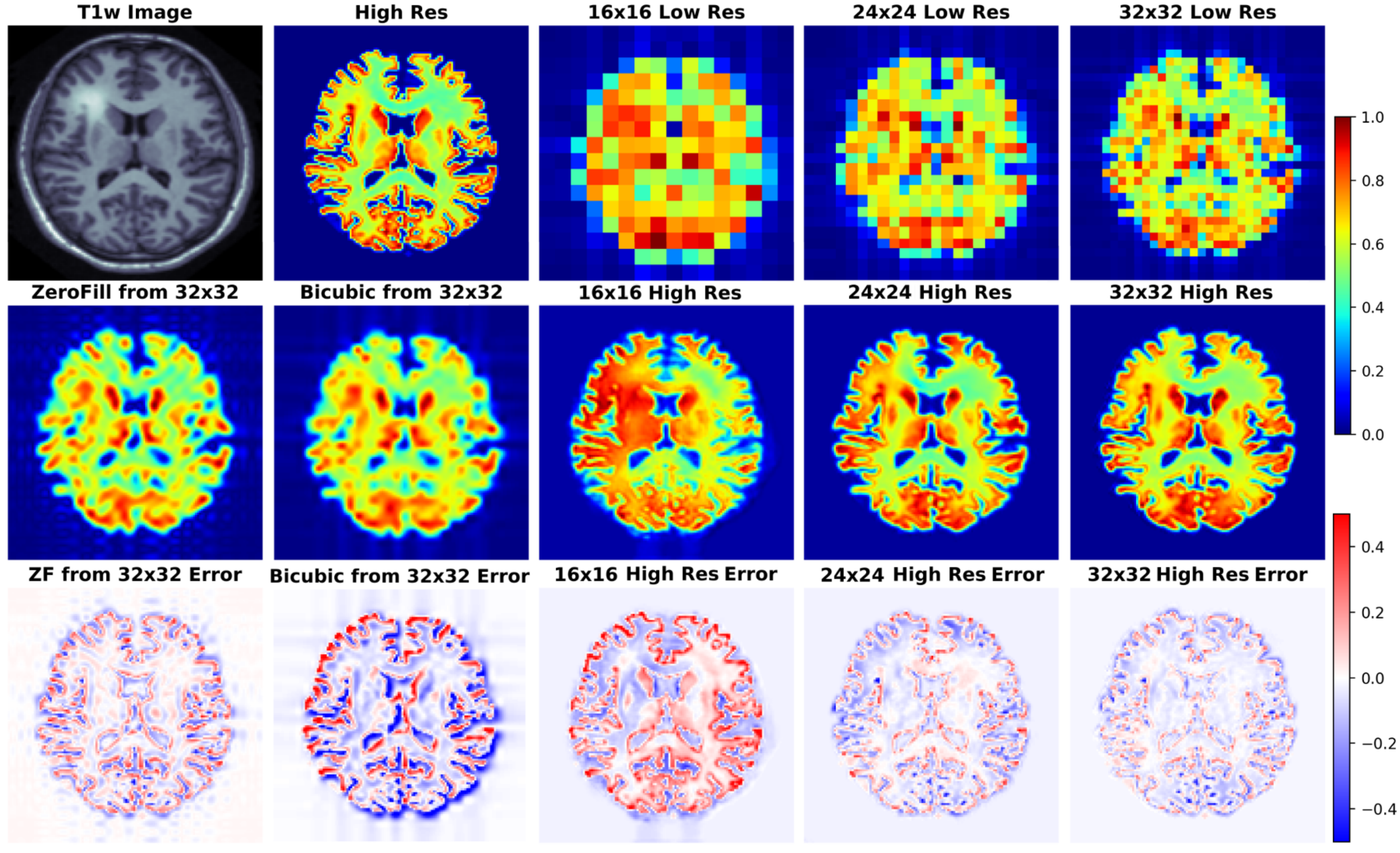}
	
	\caption{\small The results of the three D-UNets are shown for an example test subject. The augmented T1w image is used in conjunction with the three low resolution images as inputs for the three D-UNets. The reconstructed HRSI (16x16 High Res, 24x24 High Res, and 32x32 High Res) are shown below their respective low resolution images. In addition, zero-filling and bicubic interpolation were applied to the 32x32 LRSI to produce 128x128 interpolated images. Error maps are produced by subtracting the reconstructed images and the ground truth high resolution image (High Res). The 16x16 High Res displays much more error than the 24x24 and 32x32 High Res images. This is mostly due to better local signal refinement at the location of the lesion for the 24x24 and 32x32 reconstructions.}
	\label{fig:comp}
\end{figure}

\begin{figure}
	\center
	\hspace*{0cm}
	\includegraphics[scale=0.4]{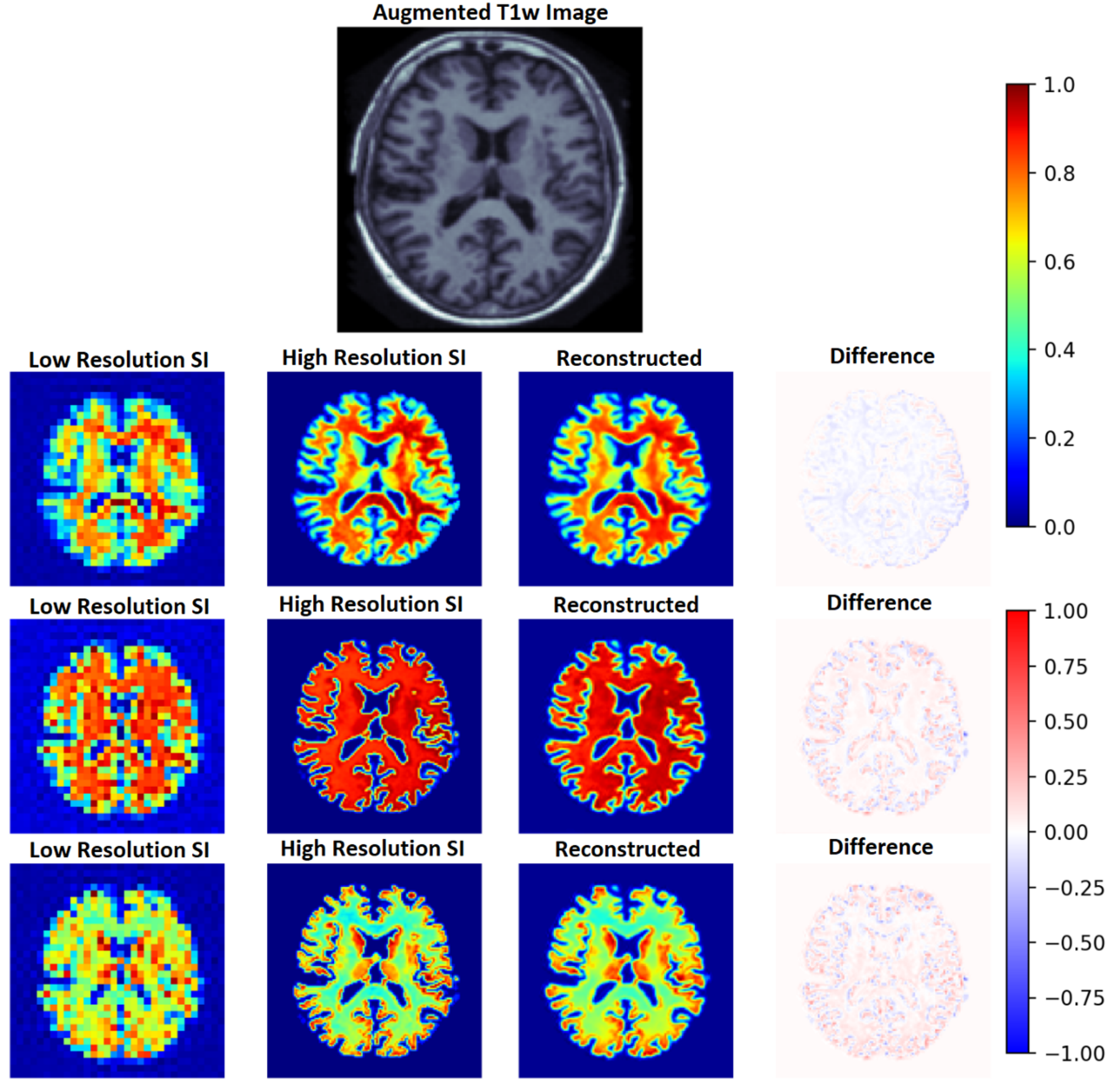}
	
	\caption{\small From one augmented T1w image, the generator is capable of producing multiple ground truth high resolution (High Resolution SI) images and low resolution (Low Resolution SI) images, a small sample of which are shown. In this example, the top row shows images where metabolite signal is higher in WM. In the middle row, the metabolite signal is equal in WM and GM, whereas the metabolite signal is higher in the GM in the bottom row. Since a single input T1w image can produce many augmented T1w images, the generator allows for an exponentially large number of unique training data. The reconstruction for each $aT1w$ image and LRSI is performed with the 32x32 D-UNet to yield the reconstructed HRSI images (Reconstructed). The difference maps are produced by subtracting the reconstructed and ground truth images.}
	\label{fig:comp2}
\end{figure}

\subsection*{Test Set Results}
Figure \ref{fig:comp} displays the results from the three different D-UNet reconstructions, as well as the results of the standard zero-filling and bicubic interpolation methods. In order to provide a more stringent comparison, both zero-filling and bicubic interpolation were applied to the 32x32 low resolution metabolite maps instead of the lower resolution 16x16 or 24x24 metabolite maps. All of the D-UNet reconstructions are able to determine the abnormally high signal from the lesion shown in the T1w image. While zero-filling outperforms both bicubic interpolation and the 16x16 D-UNet, both the 24x24 and 32x32 D-UNets yield better results than zero-filling.  

To demonstrate the capability of the SI generator, Fig. \ref{fig:comp2} shows a sample of the possible images produced from the same $aT1w$ image. The $Recon_{32x32}$ images are also shown, as well as difference maps between the HRSI and $Recon_{32x32}$. It is clear that the SI generator is capable of producing a wide variety of $SI$ images that mimic biochemicals that are more prominent in GM, more prominent in WM, or equally prominent in both tissue types.  

In addition, a quantitative comparison between these methods is shown in Table \ref{table:MSE}. Noise level was varied to determine the effect of noise on the super-resolution methods. Low noise level, medium noise level, and high noise level were classified as 2-5\%, 15-20\%, and 30-40\% of the maximum signal intensity, respectively. From Table \ref{table:MSE}, the 32x32 D-UNet demonstrated the best performance at every noise level. At medium noise levels, the 24x24 D-UNet outperformed zero-filling, and at high noise levels both the 16x16 D-UNet and 24x24 D-UNet outperformed both zero-filling and bicubic interpolation.  

\subsection*{Spectral Reconstruction Results }
\begin{figure}
	\center
	\hspace*{-1.0cm}
	\includegraphics[scale=0.33]{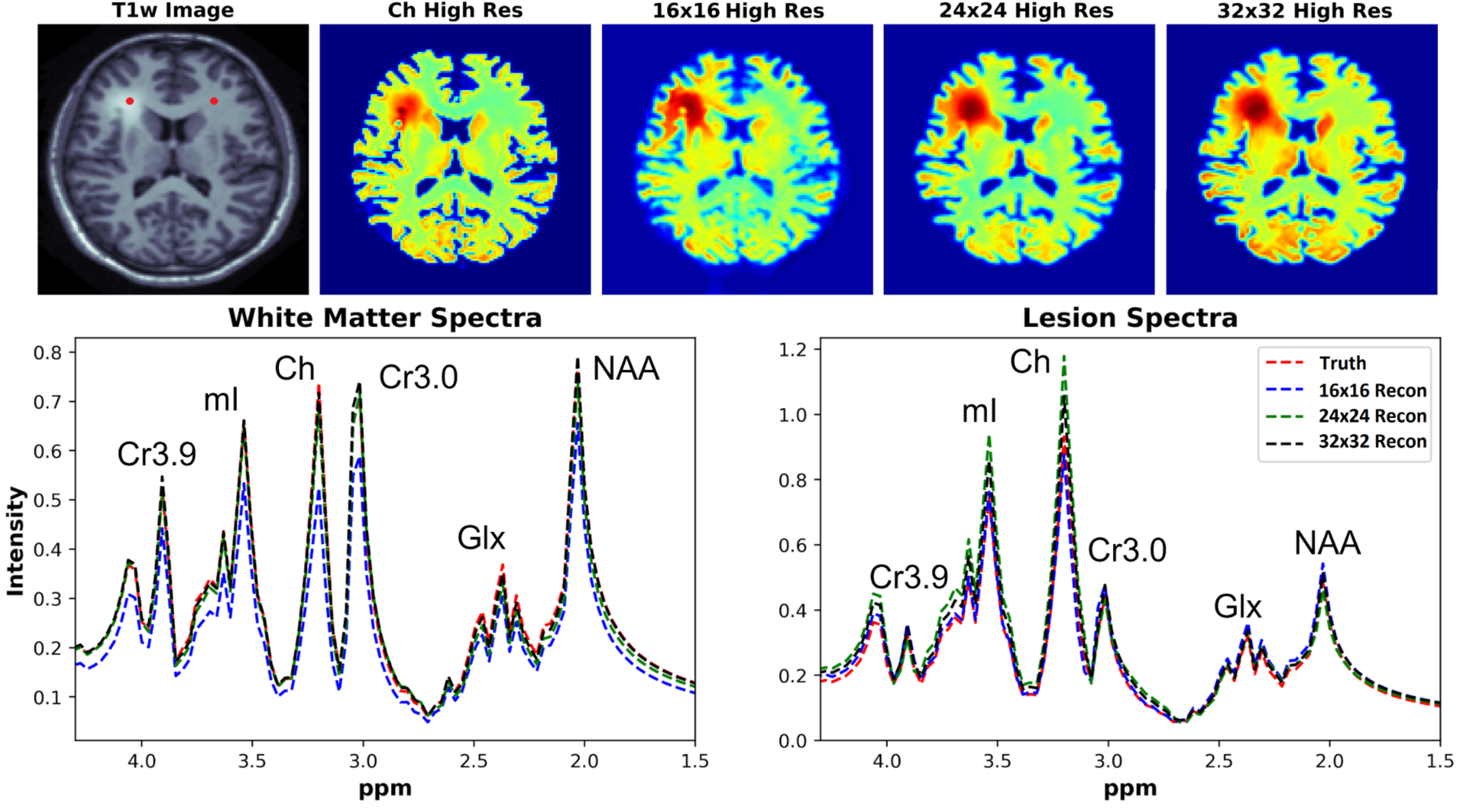}
	\caption{\small Magnitude spectral reconstructions using the three D-UNets are shown for two voxels. The voxel locations for the white matter and lesion spectra are displayed in the T1w image as red points. The spectra generated from the ground truth (red), the 16x16 (blue), 24x24 (green), and 32x32 D-UNets (black) for the 1.5-4.3ppm range are displayed. For spatial comparison, the choline metabolite maps for each method are also shown. All metabolite maps are scaled from 0 to 1. The 24x24 and 32x32 D-UNet reconstructions over-estimate the amount of choline in the lesion. However, the 16x16 reconstruction under-estimates the amount of metabolite signal in the healthy white matter region. }
	\label{fig:spec}
\end{figure}

The ability of the D-UNets to reconstruct spectra at high spatial resolutions are highlighted in Fig. \ref{fig:spec}. The 32x32 D-UNet reconstructs the lesion and contra-lateral white matter spectra reliably. In contrast, the 16x16 D-UNet underestimates the white matter spectrum. The 24x24 D-UNet performs very similarly to the 32x32 D-UNet, however it overestimates the Ch and mI signals in the lesion spectrum by roughly 20\%. Overall, the mean squared error for the healthy white matter spectrum was 0.34, 0.030, and 0.0085 for the 16x16 D-UNet, 24x24 D-UNet, and 32x32 D-UNet, respectively. For the lesion spectrum, the mean squared error was 0.051, 0.36, and 0.13 for the 16x16 D-UNet, 24x24 D-UNet, and 32x32 D-UNet, respectively. From a quantitative standpoint, all three D-UNets would be able to determine the abnormally elevated Ch, as demonstrated from the metabolite maps.
   
\subsection*{In Vivo Results }
\begin{figure}
	\center
	\hspace*{-0.75cm}
	\includegraphics[scale=0.34]{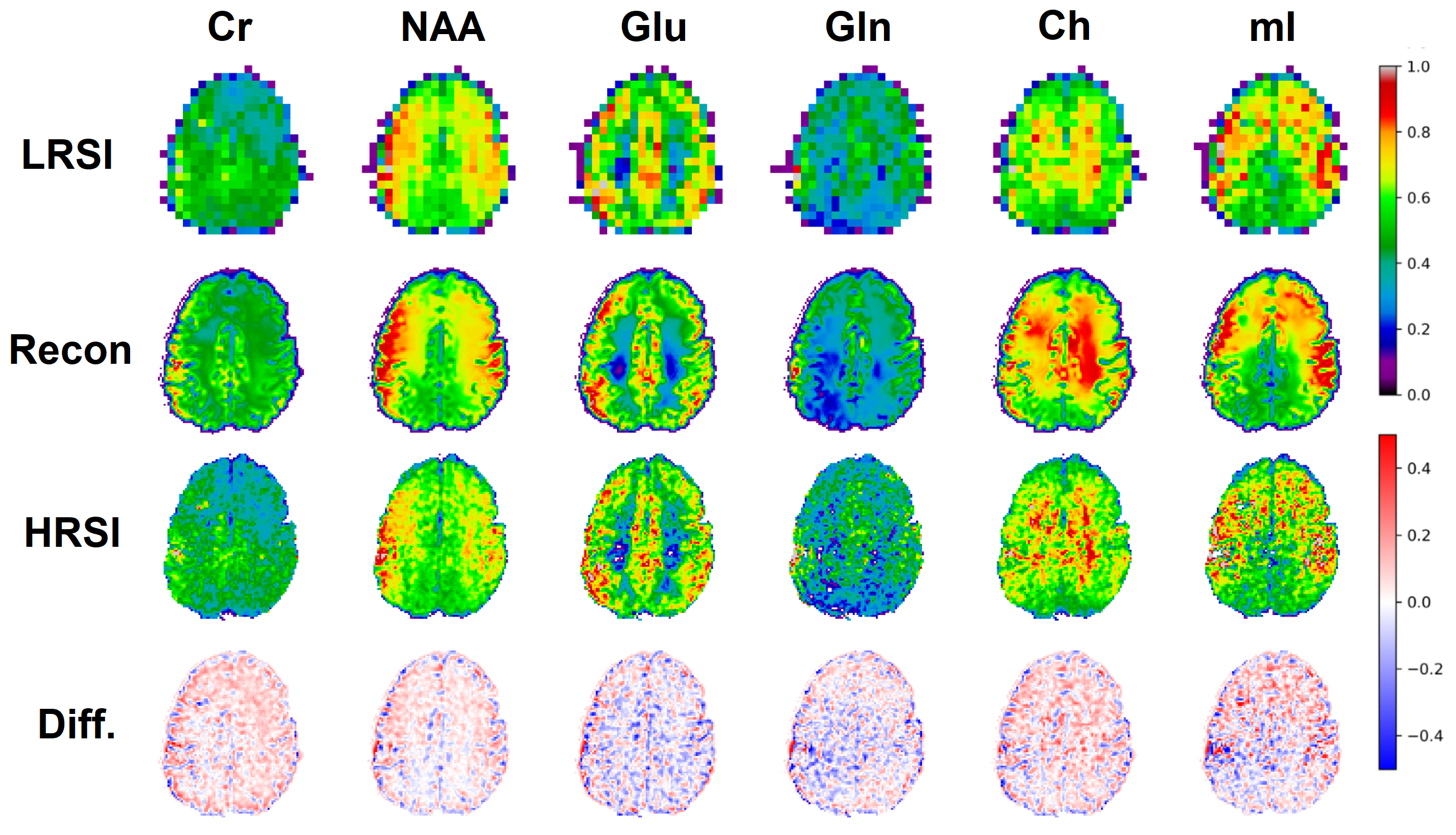}
	\caption{\small An \textit{in vivo} example of a healthy volunteer is used to demonstrate the potential application for the D-UNet. The experimental high resolution SI (HRSI) data was acquired at 128x128 resolution using an accelerated acquisition protocol\cite{hangel2016ultra}. This data was then down-sampled to produce 32x32 low resolution SI (LRSI) metabolite maps for Cr, NAA, Glu, Gln, Ch, and mI. Together with the T1w image, the low resolution metabolite images were used to reconstruct high resolution spectroscopic images (Recon) using the 32x32 D-UNet model. The difference maps between the Recon and HRSI images (Diff) are also shown. }
	\label{fig:invivo}
\end{figure}

\begin{figure}
	\center
	\hspace*{-0.75cm}
	\includegraphics[scale=0.34]{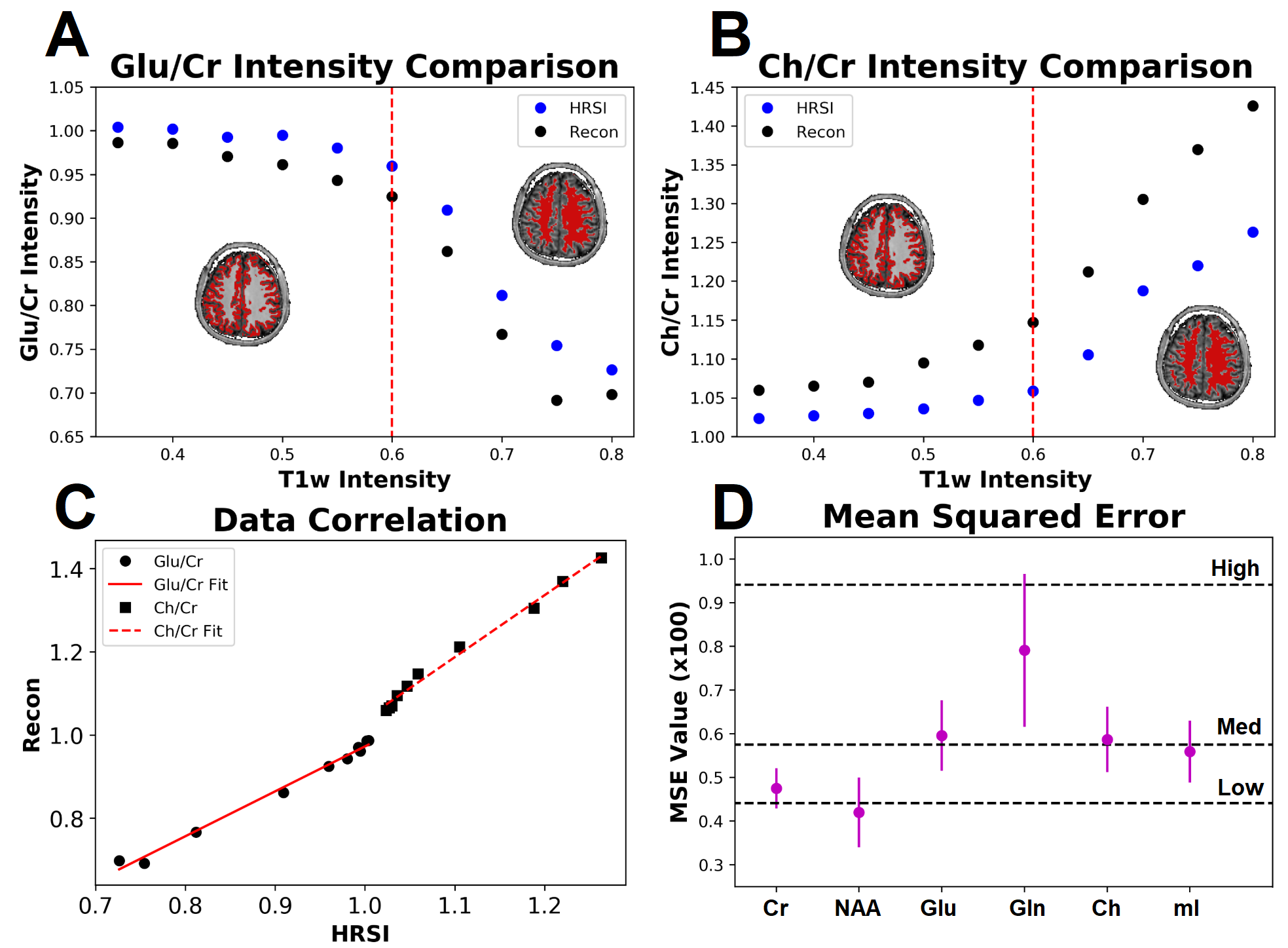}
	\caption{\small The Glu/Cr (A) and Ch/Cr (B) signals averaged over ten subjects are shown for the experimentally acquired images (HRSI) and the images reconstructed from low resolution 32x32 images (Recon). The signals are shown as a function of the T1w intensity, which is representative of the gray and white matter content of the voxel. The red dotted line represents the point at which the gray matter content equals the white matter content in a voxel. The correlation between the HRSI and Recon values are plotted (C) with linear fits. For both Glu/Cr and Ch/Cr, the r$^2$ values of the fits are above 0.99. Finally, the mean squared error for the ten subjects calculated between the HRSI and Recon for each metabolite map (D) is displayed. The dotted black lines reflect the MSE from the testing set for different noise values (low, medium, and high).   }
	\label{fig:MSE}
\end{figure}

The ability of the 32x32 D-UNet to reconstruct the LRSI of Cr, NAA, Glu, Gln, Ch, and mI is shown in Fig. \ref{fig:invivo}. This figure shows the reconstructed images, experimental HRSI, and difference maps between the two for each metabolite for one healthy volunteer. All reconstructed images retain the metabolite signals from the low resolution maps, and also show regional changes similar to the HRSI. For example, Glu is more concentrated in the GM and less concentrated in the WM, which is a well known regional difference in the brain\cite{pouwels1998regional}. Another well known regional difference is that Ch is more concentrated in WM regions, which is apparent in both the reconstructed and experimental images. From a quantitative standpoint, the average MSE values over the ten volunteers for Cr, NAA, Glu, Gln, Ch, and mI were 0.0048, 0.0042, 0.0060, 0.0079, and 0.0059, respectively. These errors are displayed in Fig. \ref{fig:MSE}D and plotted against the average MSE values obtained for the testing set using different noise levels (low, medium, high). It is clear that the MSE values are in most cases comparable to simulated test images with 2-20\% noise, with the exception of Gln which is most comparable to test images with 35\% noise. 

Figure \ref{fig:MSE} also shows the Glu/Cr and Ch/Cr ratios as a function of the T1w intensity averaged over the ten volunteers. The ratios are taken after normalization of the metabolites as part of the super-resolution reconstruction, which is why Ch/Cr appears larger than Glu/Cr in the figure. The trend shows that with higher WM content, Glu/Cr decreases while Ch/Cr increases. The correlation between the experimental HRSI and Recon results are shown in Fig. \ref{fig:MSE}C. Quantitatively, both Glu/Cr and Ch/Cr ratios have high squared correlation coefficients, $r^2$ $>$ 0.99. This highlights the fact that important biological relationships are preserved in the reconstructed images.      

\section*{Discussion}
Although SI provides invaluable information regarding the biomolecular processes of tissues \textit{in vivo}, experimental limitations have greatly hindered the integration of this method into standard clinical protocols. This study demonstrates a technique capable of overcoming one of the greatest challenges in SI, which is poor spatial resolution. By utilizing a deep learning framework, it is shown in Figures \ref{fig:comp}-\ref{fig:invivo} that high resolution spectroscopic images can be produced from the combination of low resolution spectroscopic images and T1w images. In addition, as seen in Fig. \ref{fig:spec}, it is possible to reconstruct spectra at higher spatial resolutions. The reconstruction method also preserves important regional metabolic differences and shows low errors for \textit{in vivo} reconstructions, as shown in Fig. \ref{fig:MSE}. This deep learning super-resolution method was compared to both zero-filling and bicubic interpolation, and proved to be better than these methods for all noise levels.  

Deep learning requires large datasets, which are not readily available for SI. Unfortunately, there is also a lack of ground truth for high resolution spectroscopic imaging due to the fact that experimental results may contain chemical shift displacement artifacts, B$_0$ inhomogeneity issues, partial volume effects, low signal to noise ratios, water contamination, or other forms of signal contamination. It is also prohibitively long to scan at high resolution (128x128) without using several acceleration methods, making a ground truth impossible to obtain from the human brain. Therefore, an SI generator was developed to simulate training and testing data from a publicly available dataset. By including various probabilistic transformations, such as contrast variations, metabolic signal changes, and FOV variations, the SI generator was capable of providing a diverse and large dataset for the training of the three D-UNets. These data may not be entirely realistic, and this generator must be validated more rigorously in the future. For this study, the dataset does seem to be representative of real acquisitions, as seen from the \textit{in vivo} results. 

The $Recon_{32x32}$ and HRSI experimental images are very similar, as seen from Figs. \ref{fig:invivo} and \ref{fig:MSE}. The reconstructed images show better resemblance to the anatomical T1w images, including cerebral spinal fluid localization. However, both the $Recon_{32x32}$ and HRSI experimental images provide similar quantitative results, as seen in Fig. \ref{fig:MSE}. Theoretically, the $Recon_{32x32}$ images would require $\frac{1}{16}th$ to $\frac{1}{4}th$ the time to acquire, depending on the acceleration methods implemented. Therefore, it is important to note that aside from super-resolution, the D-UNet may also be used as a means to accelerate a spectroscopic imaging protocol in the future. Additionally, the reconstructed \textit{in vivo} images are denoised while retaining essential metabolic information for different tissues of the brain, which may be desirable for certain applications. While the simulated and \textit{in vivo} data demonstrate that the reconstruction method is accurate, one of the main disadvantages of this work is that it has not been validated \textit{in vitro}. This is due to the fact that a high resolution SI phantom similar to the human brain is not available. Since the D-UNet model is trained using \textit{in vivo} anatomy, it is not capable of reconstructing high resolution images from unrealistic geometries. Therefore, future work will focus on the development of a realistic, high resolution SI phantom for validation.

Even though the D-UNets outperformed zero-filling and bicubic interpolation, these models may not be perfect for HRSI reconstruction primarily due to experimental imperfections. As seen from Table \ref{table:MSE}, error increases as a function of noise. Intuitively, chemicals that are found in the body at lower concentration may have larger reconstruction errors than chemicals with higher SNR, which is also supported by \textit{in vivo} results where the Gln reconstructed images have higher error than the other metabolite images. Therefore, prediction accuracy is limited by the quality of the original LRSI.  

Of course, the original resolution of the experimental SI plays a large role in the reconstruction process. While 24x24 and 32x32 matrices provide relatively accurate high resolution reconstructions, the 16x16 resolution does not perform as well. This suggests that there is a lower bound necessary to accurately upscale high resolution SI. This might be true for other super-resolution techniques\cite{jain2017patch}, so a more thorough comparison between this deep learning method and other methods may aid in identifying this lower bound. Furthermore, results may be biased by the quantitative methods implemented to produce the LRSI before the super-resolution process is performed. This bias could be removed in the future by developing a deep learning based approach to metabolite quantitation\cite{hatami2018magnetic}, however it may be worthwhile to explore the differences between common one dimensional spectral quantitation programs, such as LCModel\cite{provencher1993estimation} or TARQUIN\cite{wilson2011constrained}, on the upscaling process.  

The deep learning method presented in this study may be useful for other super-resolution transformations in the field of medical imaging. For example, the same principles discussed in this work apply to positron emission tomography (PET)\cite{kennedy2006super}. It is well known that the radioactive tracer is more prominent in certain tissues and lesions, and positrons from this tracer travel some distance before annihilating to produce the PET signal. The distance between the source and the annihilation can be thought of as a partial volume effect. This model can potentially be used to learn how to remove this partial volume effect artifact, and this would be applicable for CT-PET or MR-PET acquisitions. Ultimately, this deep learning model allows for the acquisition of high quality images without increasing the scan time or improving the hardware of the imaging system.

\section*{Conclusion}
The D-UNet model presented in this study allows for the reconstruction of accurate super-resolution magnetic resonance spectroscopic images from the human brain. Utilizing this method, a low resolution chemical map produced from any SI protocol can be transformed together with the T1w image to produce a high resolution chemical map. This method demonstrates better accuracy than typical zero-filling and bicubic interpolation methods, and also demonstrates the capability to reconstruct spectra faithfully to a higher spatial resolution. This model can be utilized for denoising, scan acceleration, and improved tissue delineation after further \textit{in vitro} and \textit{in vivo} validation.

\section*{Acknowledgments}
The authors would like to acknowledge the support of NIH/NCI (1R01CA154747-01), the open source MRI data provided by the OASIS project (funded by grants P50 AG05681, P01 AG03991, R01 AG021910, P20 MH071616, and U24 RR021382), and the Austrian Science Fund (FWF): KLI 646 and P 30701.  

\section*{Author Contributions Statement}

ZI and SJ conceived the experiments, DN designed the deep learning architecture, ZI and DN conducted the deep learning experiments, GH, SM, WB acquired and processed the \textit{in vivo} data, ZI and SJ analyzed the results. All authors reviewed the manuscript. 

\section*{Additional Information}
The authors do not have any competing financial and/or non-financial interests in relation to the work described.


\begin{thebibliography}{10}
\expandafter\ifx\csname url\endcsname\relax
  \def\url#1{\texttt{#1}}\fi
\expandafter\ifx\csname urlprefix\endcsname\relax\def\urlprefix{URL }\fi
\expandafter\ifx\csname doiprefix\endcsname\relax\def\doiprefix{DOI }\fi
\providecommand{\bibinfo}[2]{#2}
\providecommand{\eprint}[2][]{\url{#2}}

\bibitem{brown1982nmr}
\bibinfo{author}{Brown, T.}, \bibinfo{author}{Kincaid, B.} \&
  \bibinfo{author}{Ugurbil, K.}
\newblock \bibinfo{title}{Nmr chemical shift imaging in three dimensions}.
\newblock \emph{\bibinfo{journal}{Proceedings of the National Academy of
  Sciences}} \textbf{\bibinfo{volume}{79}}, \bibinfo{pages}{3523--3526}
  (\bibinfo{year}{1982}).

\bibitem{soares2009magnetic}
\bibinfo{author}{Soares, D.} \& \bibinfo{author}{Law, M.}
\newblock \bibinfo{title}{Magnetic resonance spectroscopy of the brain: review
  of metabolites and clinical applications}.
\newblock \emph{\bibinfo{journal}{Clinical radiology}}
  \textbf{\bibinfo{volume}{64}}, \bibinfo{pages}{12--21}
  (\bibinfo{year}{2009}).

\bibitem{govindaraju2000proton}
\bibinfo{author}{Govindaraju, V.}, \bibinfo{author}{Young, K.} \&
  \bibinfo{author}{Maudsley, A.~A.}
\newblock \bibinfo{title}{Proton nmr chemical shifts and coupling constants for
  brain metabolites}.
\newblock \emph{\bibinfo{journal}{NMR in Biomedicine}}
  \textbf{\bibinfo{volume}{13}}, \bibinfo{pages}{129--153}
  (\bibinfo{year}{2000}).

\bibitem{ramadan2013glutamate}
\bibinfo{author}{Ramadan, S.}, \bibinfo{author}{Lin, A.} \&
  \bibinfo{author}{Stanwell, P.}
\newblock \bibinfo{title}{Glutamate and glutamine: a review of in vivo mrs in
  the human brain}.
\newblock \emph{\bibinfo{journal}{NMR in Biomedicine}}
  \textbf{\bibinfo{volume}{26}}, \bibinfo{pages}{1630--1646}
  (\bibinfo{year}{2013}).

\bibitem{mansfield1984spatial}
\bibinfo{author}{Mansfield, P.}
\newblock \bibinfo{title}{Spatial mapping of the chemical shift in nmr}.
\newblock \emph{\bibinfo{journal}{Magnetic Resonance in Medicine}}
  \textbf{\bibinfo{volume}{1}}, \bibinfo{pages}{370--386}
  (\bibinfo{year}{1984}).

\bibitem{posse1995high}
\bibinfo{author}{Posse, S.}, \bibinfo{author}{Tedeschi, G.},
  \bibinfo{author}{Risinger, R.}, \bibinfo{author}{Ogg, R.} \&
  \bibinfo{author}{Bihan, D.~L.}
\newblock \bibinfo{title}{High speed 1h spectroscopic imaging in human brain by
  echo planar spatial-spectral encoding}.
\newblock \emph{\bibinfo{journal}{Magnetic Resonance in Medicine}}
  \textbf{\bibinfo{volume}{33}}, \bibinfo{pages}{34--40}
  (\bibinfo{year}{1995}).

\bibitem{adalsteinsson1998volumetric}
\bibinfo{author}{Adalsteinsson, E.} \emph{et~al.}
\newblock \bibinfo{title}{Volumetric spectroscopic imaging with spiral-based
  k-space trajectories}.
\newblock \emph{\bibinfo{journal}{Magnetic resonance in medicine}}
  \textbf{\bibinfo{volume}{39}}, \bibinfo{pages}{889--898}
  (\bibinfo{year}{1998}).

\bibitem{furuyama2012spectroscopic}
\bibinfo{author}{Furuyama, J.~K.}, \bibinfo{author}{Wilson, N.~E.} \&
  \bibinfo{author}{Thomas, M.~A.}
\newblock \bibinfo{title}{Spectroscopic imaging using concentrically circular
  echo-planar trajectories in vivo}.
\newblock \emph{\bibinfo{journal}{Magnetic resonance in medicine}}
  \textbf{\bibinfo{volume}{67}}, \bibinfo{pages}{1515--1522}
  (\bibinfo{year}{2012}).

\bibitem{schirda2009rosette}
\bibinfo{author}{Schirda, C.~V.}, \bibinfo{author}{Tanase, C.} \&
  \bibinfo{author}{Boada, F.~E.}
\newblock \bibinfo{title}{Rosette spectroscopic imaging: Optimal parameters for
  alias-free, high sensitivity spectroscopic imaging}.
\newblock \emph{\bibinfo{journal}{Journal of Magnetic Resonance Imaging}}
  \textbf{\bibinfo{volume}{29}}, \bibinfo{pages}{1375--1385}
  (\bibinfo{year}{2009}).

\bibitem{pruessmann1999sense}
\bibinfo{author}{Pruessmann, K.~P.}, \bibinfo{author}{Weiger, M.},
  \bibinfo{author}{Scheidegger, M.~B.}, \bibinfo{author}{Boesiger, P.}
  \emph{et~al.}
\newblock \bibinfo{title}{Sense: sensitivity encoding for fast mri}.
\newblock \emph{\bibinfo{journal}{Magnetic Resonance in Medicine}}
  \textbf{\bibinfo{volume}{42}}, \bibinfo{pages}{952--962}
  (\bibinfo{year}{1999}).

\bibitem{dydak2001sensitivity}
\bibinfo{author}{Dydak, U.}, \bibinfo{author}{Weiger, M.},
  \bibinfo{author}{Pruessmann, K.~P.}, \bibinfo{author}{Meier, D.} \&
  \bibinfo{author}{Boesiger, P.}
\newblock \bibinfo{title}{Sensitivity-encoded spectroscopic imaging}.
\newblock \emph{\bibinfo{journal}{Magnetic Resonance in Medicine}}
  \textbf{\bibinfo{volume}{46}}, \bibinfo{pages}{713--722}
  (\bibinfo{year}{2001}).

\bibitem{griswold2002generalized}
\bibinfo{author}{Griswold, M.~A.} \emph{et~al.}
\newblock \bibinfo{title}{Generalized autocalibrating partially parallel
  acquisitions (grappa)}.
\newblock \emph{\bibinfo{journal}{Magnetic Resonance in Medicine}}
  \textbf{\bibinfo{volume}{47}}, \bibinfo{pages}{1202--1210}
  (\bibinfo{year}{2002}).

\bibitem{otazo2007accelerated}
\bibinfo{author}{Otazo, R.}, \bibinfo{author}{Tsai, S.-Y.},
  \bibinfo{author}{Lin, F.-H.} \& \bibinfo{author}{Posse, S.}
\newblock \bibinfo{title}{Accelerated short-te 3d proton echo-planar
  spectroscopic imaging using 2d-sense with a 32-channel array coil}.
\newblock \emph{\bibinfo{journal}{Magnetic Resonance in Medicine}}
  \textbf{\bibinfo{volume}{58}}, \bibinfo{pages}{1107--1116}
  (\bibinfo{year}{2007}).

\bibitem{strasser20172+}
\bibinfo{author}{Strasser, B.} \emph{et~al.}
\newblock \bibinfo{title}{(2+ 1) d-caipirinha accelerated mr spectroscopic
  imaging of the brain at 7t}.
\newblock \emph{\bibinfo{journal}{Magnetic resonance in medicine}}
  \textbf{\bibinfo{volume}{78}}, \bibinfo{pages}{429--440}
  (\bibinfo{year}{2017}).

\bibitem{wilson2015accelerated}
\bibinfo{author}{Wilson, N.~E.}, \bibinfo{author}{Iqbal, Z.},
  \bibinfo{author}{Burns, B.~L.}, \bibinfo{author}{Keller, M.} \&
  \bibinfo{author}{Thomas, M.~A.}
\newblock \bibinfo{title}{Accelerated five-dimensional echo planar j-resolved
  spectroscopic imaging: Implementation and pilot validation in human brain}.
\newblock \emph{\bibinfo{journal}{Magnetic Resonance in Medicine}}
  \textbf{\bibinfo{volume}{75}}, \bibinfo{pages}{42--51}
  (\bibinfo{year}{2016}).

\bibitem{iqbal20163d}
\bibinfo{author}{Iqbal, Z.}, \bibinfo{author}{Wilson, N.~E.} \&
  \bibinfo{author}{Thomas, M.~A.}
\newblock \bibinfo{title}{3d spatially encoded and accelerated te-averaged echo
  planar spectroscopic imaging in healthy human brain}.
\newblock \emph{\bibinfo{journal}{NMR in Biomedicine}}
  \textbf{\bibinfo{volume}{29}}, \bibinfo{pages}{329--339}
  (\bibinfo{year}{2016}).

\bibitem{posse2013mr}
\bibinfo{author}{Posse, S.}, \bibinfo{author}{Otazo, R.},
  \bibinfo{author}{Dager, S.~R.} \& \bibinfo{author}{Alger, J.}
\newblock \bibinfo{title}{Mr spectroscopic imaging: principles and recent
  advances}.
\newblock \emph{\bibinfo{journal}{Journal of Magnetic Resonance Imaging}}
  \textbf{\bibinfo{volume}{37}}, \bibinfo{pages}{1301--1325}
  (\bibinfo{year}{2013}).

\bibitem{lam2014subspace}
\bibinfo{author}{Lam, F.} \& \bibinfo{author}{Liang, Z.-P.}
\newblock \bibinfo{title}{A subspace approach to high-resolution spectroscopic
  imaging}.
\newblock \emph{\bibinfo{journal}{Magnetic resonance in medicine}}
  \textbf{\bibinfo{volume}{71}}, \bibinfo{pages}{1349--1357}
  (\bibinfo{year}{2014}).

\bibitem{hingerl2017density}
\bibinfo{author}{Hingerl, L.} \emph{et~al.}
\newblock \bibinfo{title}{Density-weighted concentric circle trajectories for
  high resolution brain magnetic resonance spectroscopic imaging at 7t}.
\newblock \emph{\bibinfo{journal}{Magnetic resonance in medicine}}
  (\bibinfo{year}{2017}).

\bibitem{hangel2016ultra}
\bibinfo{author}{Hangel, G.} \emph{et~al.}
\newblock \bibinfo{title}{Ultra-high resolution brain metabolite mapping at 7 t
  by short-tr hadamard-encoded fid-mrsi}.
\newblock \emph{\bibinfo{journal}{Neuroimage}}  (\bibinfo{year}{2016}).

\bibitem{jain2017patch}
\bibinfo{author}{Jain, S.} \emph{et~al.}
\newblock \bibinfo{title}{Patch-based super-resolution of mr spectroscopic
  images: application to multiple sclerosis}.
\newblock \emph{\bibinfo{journal}{Frontiers in neuroscience}}
  \textbf{\bibinfo{volume}{11}}, \bibinfo{pages}{13} (\bibinfo{year}{2017}).

\bibitem{haldar2008anatomically}
\bibinfo{author}{Haldar, J.~P.}, \bibinfo{author}{Hernando, D.},
  \bibinfo{author}{Song, S.-K.} \& \bibinfo{author}{Liang, Z.-P.}
\newblock \bibinfo{title}{Anatomically constrained reconstruction from noisy
  data}.
\newblock \emph{\bibinfo{journal}{Magnetic Resonance in Medicine}}
  \textbf{\bibinfo{volume}{59}}, \bibinfo{pages}{810--818}
  (\bibinfo{year}{2008}).

\bibitem{hu1988slim}
\bibinfo{author}{Hu, X.}, \bibinfo{author}{Levin, D.~N.},
  \bibinfo{author}{Lauterbur, P.~C.} \& \bibinfo{author}{Spraggins, T.}
\newblock \bibinfo{title}{Slim: Spectral localization by imaging}.
\newblock \emph{\bibinfo{journal}{Magnetic resonance in medicine}}
  \textbf{\bibinfo{volume}{8}}, \bibinfo{pages}{314--322}
  (\bibinfo{year}{1988}).

\bibitem{cengiz2017super}
\bibinfo{author}{Cengiz, S.}, \bibinfo{author}{Valdes-Hernandez, M. d.~C.} \&
  \bibinfo{author}{Ozturk-Isik, E.}
\newblock \bibinfo{title}{Super resolution convolutional neural networks for
  increasing spatial resolution of $ $\^{}$\{$1$\}$ $ $ h magnetic resonance
  spectroscopic imaging}.
\newblock In \emph{\bibinfo{booktitle}{Annual Conference on Medical Image
  Understanding and Analysis}}, \bibinfo{pages}{641--650}
  (\bibinfo{organization}{Springer}, \bibinfo{year}{2017}).

\bibitem{jacob2007improved}
\bibinfo{author}{Jacob, M.}, \bibinfo{author}{Zhu, X.}, \bibinfo{author}{Ebel,
  A.}, \bibinfo{author}{Schuff, N.} \& \bibinfo{author}{Liang, Z.-P.}
\newblock \bibinfo{title}{Improved model-based magnetic resonance spectroscopic
  imaging}.
\newblock \emph{\bibinfo{journal}{IEEE transactions on medical imaging}}
  \textbf{\bibinfo{volume}{26}}, \bibinfo{pages}{1305--1318}
  (\bibinfo{year}{2007}).

\bibitem{liang1991generalized}
\bibinfo{author}{Liang, Z.-P.} \& \bibinfo{author}{Lauterbur, P.~C.}
\newblock \bibinfo{title}{A generalized series approach to mr spectroscopic
  imaging}.
\newblock \emph{\bibinfo{journal}{IEEE Transactions on Medical imaging}}
  \textbf{\bibinfo{volume}{10}}, \bibinfo{pages}{132--137}
  (\bibinfo{year}{1991}).

\bibitem{kasten2016magnetic}
\bibinfo{author}{Kasten, J.}, \bibinfo{author}{Klauser, A.},
  \bibinfo{author}{Lazeyras, F.} \& \bibinfo{author}{Van De~Ville, D.}
\newblock \bibinfo{title}{Magnetic resonance spectroscopic imaging at
  superresolution: overview and perspectives}.
\newblock \emph{\bibinfo{journal}{Journal of Magnetic Resonance}}
  \textbf{\bibinfo{volume}{263}}, \bibinfo{pages}{193--208}
  (\bibinfo{year}{2016}).

\bibitem{lecun1989backpropagation}
\bibinfo{author}{LeCun, Y.} \emph{et~al.}
\newblock \bibinfo{title}{Backpropagation applied to handwritten zip code
  recognition}.
\newblock \emph{\bibinfo{journal}{Neural computation}}
  \textbf{\bibinfo{volume}{1}}, \bibinfo{pages}{541--551}
  (\bibinfo{year}{1989}).

\bibitem{lecun2015deep}
\bibinfo{author}{LeCun, Y.}, \bibinfo{author}{Bengio, Y.} \&
  \bibinfo{author}{Hinton, G.}
\newblock \bibinfo{title}{Deep learning}.
\newblock \emph{\bibinfo{journal}{nature}} \textbf{\bibinfo{volume}{521}},
  \bibinfo{pages}{436} (\bibinfo{year}{2015}).

\bibitem{krizhevsky2012imagenet}
\bibinfo{author}{Krizhevsky, A.}, \bibinfo{author}{Sutskever, I.} \&
  \bibinfo{author}{Hinton, G.~E.}
\newblock \bibinfo{title}{Imagenet classification with deep convolutional
  neural networks}.
\newblock In \emph{\bibinfo{booktitle}{Advances in neural information
  processing systems}}, \bibinfo{pages}{1097--1105} (\bibinfo{year}{2012}).

\bibitem{ronneberger2015u}
\bibinfo{author}{Ronneberger, O.}, \bibinfo{author}{Fischer, P.} \&
  \bibinfo{author}{Brox, T.}
\newblock \bibinfo{title}{U-net: Convolutional networks for biomedical image
  segmentation}.
\newblock In \emph{\bibinfo{booktitle}{International Conference on Medical
  image computing and computer-assisted intervention}},
  \bibinfo{pages}{234--241} (\bibinfo{organization}{Springer},
  \bibinfo{year}{2015}).

\bibitem{marcus2007open}
\bibinfo{author}{Marcus, D.~S.} \emph{et~al.}
\newblock \bibinfo{title}{Open access series of imaging studies (oasis):
  cross-sectional mri data in young, middle aged, nondemented, and demented
  older adults}.
\newblock \emph{\bibinfo{journal}{Journal of cognitive neuroscience}}
  \textbf{\bibinfo{volume}{19}}, \bibinfo{pages}{1498--1507}
  (\bibinfo{year}{2007}).

\bibitem{huang2017densely}
\bibinfo{author}{Huang, G.}, \bibinfo{author}{Liu, Z.},
  \bibinfo{author}{Weinberger, K.~Q.} \& \bibinfo{author}{van~der Maaten, L.}
\newblock \bibinfo{title}{Densely connected convolutional networks}.
\newblock In \emph{\bibinfo{booktitle}{Proceedings of the IEEE conference on
  computer vision and pattern recognition}}, vol.~\bibinfo{volume}{1},
  \bibinfo{pages}{3} (\bibinfo{year}{2017}).

\bibitem{srivastava2014dropout}
\bibinfo{author}{Srivastava, N.}, \bibinfo{author}{Hinton, G.},
  \bibinfo{author}{Krizhevsky, A.}, \bibinfo{author}{Sutskever, I.} \&
  \bibinfo{author}{Salakhutdinov, R.}
\newblock \bibinfo{title}{Dropout: A simple way to prevent neural networks from
  overfitting}.
\newblock \emph{\bibinfo{journal}{The Journal of Machine Learning Research}}
  \textbf{\bibinfo{volume}{15}}, \bibinfo{pages}{1929--1958}
  (\bibinfo{year}{2014}).

\bibitem{kingma2014adam}
\bibinfo{author}{Kingma, D.~P.} \& \bibinfo{author}{Ba, J.}
\newblock \bibinfo{title}{Adam: A method for stochastic optimization}.
\newblock \emph{\bibinfo{journal}{arXiv preprint arXiv:1412.6980}}
  (\bibinfo{year}{2014}).

\bibitem{chollet2015keras}
\bibinfo{author}{Chollet, F.} \emph{et~al.}
\newblock \bibinfo{title}{Keras}.
\newblock \emph{\bibinfo{journal}{https://github.com/fchollet/keras}}
  (\bibinfo{year}{2015}).

\bibitem{abadi2016tensorflow}
\bibinfo{author}{Abadi, M.} \emph{et~al.}
\newblock \bibinfo{title}{Tensorflow: Large-scale machine learning on
  heterogeneous distributed systems}.
\newblock \emph{\bibinfo{journal}{arXiv preprint arXiv:1603.04467}}
  (\bibinfo{year}{2016}).

\bibitem{smith1994computer}
\bibinfo{author}{Smith, S.}, \bibinfo{author}{Levante, T.},
  \bibinfo{author}{Meier, B.~H.} \& \bibinfo{author}{Ernst, R.~R.}
\newblock \bibinfo{title}{Computer simulations in magnetic resonance. an
  object-oriented programming approach}.
\newblock \emph{\bibinfo{journal}{Journal of Magnetic Resonance, Series A}}
  \textbf{\bibinfo{volume}{106}}, \bibinfo{pages}{75--105}
  (\bibinfo{year}{1994}).

\bibitem{bogner2012high}
\bibinfo{author}{Bogner, W.}, \bibinfo{author}{Gruber, S.},
  \bibinfo{author}{Trattnig, S.} \& \bibinfo{author}{Chmelik, M.}
\newblock \bibinfo{title}{High-resolution mapping of human brain metabolites by
  free induction decay 1h mrsi at 7 t}.
\newblock \emph{\bibinfo{journal}{NMR in Biomedicine}}
  \textbf{\bibinfo{volume}{25}}, \bibinfo{pages}{873--882}
  (\bibinfo{year}{2012}).

\bibitem{bilgic2014fast}
\bibinfo{author}{Bilgic, B.} \emph{et~al.}
\newblock \bibinfo{title}{Fast image reconstruction with l2-regularization}.
\newblock \emph{\bibinfo{journal}{Journal of Magnetic Resonance Imaging}}
  \textbf{\bibinfo{volume}{40}}, \bibinfo{pages}{181--191}
  (\bibinfo{year}{2014}).

\bibitem{provencher1993estimation}
\bibinfo{author}{Provencher, S.~W.}
\newblock \bibinfo{title}{Estimation of metabolite concentrations from
  localized in vivo proton nmr spectra}.
\newblock \emph{\bibinfo{journal}{Magnetic Resonance in Medicine}}
  \textbf{\bibinfo{volume}{30}}, \bibinfo{pages}{672--679}
  (\bibinfo{year}{1993}).

\bibitem{pouwels1998regional}
\bibinfo{author}{Pouwels, P.~J.} \& \bibinfo{author}{Frahm, J.}
\newblock \bibinfo{title}{Regional metabolite concentrations in human brain as
  determined by quantitative localized proton mrs}.
\newblock \emph{\bibinfo{journal}{Magnetic Resonance in Medicine}}
  \textbf{\bibinfo{volume}{39}}, \bibinfo{pages}{53--60}
  (\bibinfo{year}{1998}).

\bibitem{hatami2018magnetic}
\bibinfo{author}{Hatami, Nima}, \bibinfo{author}{Sdika, Micha{\"e}l} \& \bibinfo{author}{Ratiney, H{\'e}l{\`e}ne}
\newblock \bibinfo{title}{Magnetic Resonance Spectroscopy Quantification using Deep Learning}
\newblock \emph{\bibinfo{journal}{arXiv preprint arXiv:1806.07237}}
  (\bibinfo{year}{2004}).


\bibitem{wilson2011constrained}
\bibinfo{author}{Wilson, M.}, \bibinfo{author}{Reynolds, G.},
  \bibinfo{author}{Kauppinen, R.~A.}, \bibinfo{author}{Arvanitis, T.~N.} \&
  \bibinfo{author}{Peet, A.~C.}
\newblock \bibinfo{title}{A constrained least-squares approach to the automated
  quantitation of in vivo 1h magnetic resonance spectroscopy data}.
\newblock \emph{\bibinfo{journal}{Magnetic Resonance in Medicine}}
  \textbf{\bibinfo{volume}{65}}, \bibinfo{pages}{1--12} (\bibinfo{year}{2011}).

\bibitem{kennedy2006super}
\bibinfo{author}{Kennedy, J.~A.}, \bibinfo{author}{Israel, O.},
  \bibinfo{author}{Frenkel, A.}, \bibinfo{author}{Bar-Shalom, R.} \&
  \bibinfo{author}{Azhari, H.}
\newblock \bibinfo{title}{Super-resolution in pet imaging}.
\newblock \emph{\bibinfo{journal}{IEEE transactions on medical imaging}}
  \textbf{\bibinfo{volume}{25}}, \bibinfo{pages}{137--147}
  (\bibinfo{year}{2006}).

\end{thebibliography}

\noindent

\newpage

\section*{List of Tables}

Table \ref{table:MSE}. The mean squared error between the high resolution ground truth ($SI_{HR}$) and several methods are tabulated. These values are the total sum of the mean squared error over 169 test subjects. The 32x32 D-UNet reconstruction outperforms all of the other methods. With higher noise present in the LRSI, the 16x16 and 24x24 D-UNets outperform both zero-filling and bicubic interpolation. It is important to note that this is true even though the zero-filling and bicubic interpolation methods are applied to a 32x32 image.

\newpage

\begin{table}[!h]
\resizebox{2\textwidth}{!}{\begin{minipage}{\textwidth}
\begin{tabular}{cccc}
\toprule
Method & \multicolumn{3}{c}{Noise Level} \\ \midrule
 & Low & Medium & High \\
 Zero-Fill from 32x32 & 1.109 & 1.652 & 4.505\\
 Bicubic from 32x32 & 2.794 & 3.129 & 3.820    \\
 16x16 D-UNet & 1.863 & 2.420 & 2.761    \\
 24x24 D-UNet & 1.139 & 1.316 & 1.745      \\
 32x32 D-UNet & \textbf{0.7460} & \textbf{0.9722} & \textbf{1.599}     \\ \bottomrule
\end{tabular}
\end{minipage}}
\caption{\small The mean squared error between the high resolution ground truth ($SI_{HR}$) and several methods are tabulated. These values are the total sum of the mean squared error over 169 test subjects. The 32x32 D-UNet reconstruction outperforms all of the other methods. With higher noise present in the LRSI, the 16x16 and 24x24 D-UNets outperform both zero-filling and bicubic interpolation. It is important to note that this is true even though the zero-filling and bicubic interpolation methods are applied to a 32x32 image.}
\label{table:MSE}
\end{table}

\newpage

\section*{List of Figures}

Figure \ref{fig:flow}. The workflow for training the D-UNet model is shown. The SI generator provides a dataset consisting of an augmented T1w image, a low resolution spectroscopic image, and a ground truth high resolution spectroscopic image. Then, the network transforms the $aT1w$ and LRSI into an initial HRSI reconstruction. This reconstruction is compared to the ground truth, and the mean squared error is calculated. Utilizing this error, the model changes the weighting parameters for the features, and continues training by using a different dataset. After training on 102,000 datasets, the model weights are refined and the reconstruction errors are minimized. 

Figure \ref{fig:dunet}. The general D-UNet architecture is displayed. Each forward convolution consisted of a convolutional layer and a concatenation process. This concatenation carries over important features which can be used to make the next layer more intelligent. In addition to local concatenations, certain features were concatenated to deeper layers in the network. This allows for prior information to improve the overall reconstruction quality. In order to use the most information possible, the last convolutional layer contains all of the carried over features.

Figure \ref{fig:loss}. The loss functions for the 16x16 D-UNet (red), 24x24 D-UNet (blue), and the 32x32 D-UNet (black) are shown. All three loss functions drop significantly in the first 2 epochs, and then gradually decrease as the training continues. Overfitting is not an issue with the current training method, since all of the data are only seen by the network once. While more epochs could be used, the loss function flattens after 70 epochs, which implies that further training will yield minimal improvement.

Figure \ref{fig:comp}. The results of the three D-UNets are shown for an example test subject. The augmented T1w image is used in conjunction with the three low resolution images as inputs for the three D-UNets. The reconstructed HRSI (16x16 High Res, 24x24 High Res, and 32x32 High Res) are shown below their respective low resolution images. In addition, zero-filling and bicubic interpolation were applied to the 32x32 LRSI to produce 128x128 interpolated images. Error maps are produced by subtracting the reconstructed images and the ground truth high resolution image (High Res). The 16x16 High Res displays much more error than the 24x24 and 32x32 High Res images. This is mostly due to better local signal refinement at the location of the lesion for the 24x24 and 32x32 reconstructions.

Figure \ref{fig:comp2}. From one augmented T1w image, the generator is capable of producing multiple ground truth high resolution (High Resolution SI) images and low resolution (Low Resolution SI) images, a small sample of which are shown. In this example, the top row shows images where metabolite signal is higher in WM. In the middle row, the metabolite signal is equal in WM and GM, whereas the metabolite signal is higher in the GM in the bottom row. Since a single input T1w image can produce many augmented T1w images, the generator allows for an exponentially large number of unique training data. The reconstruction for each $aT1w$ image and LRSI is performed with the 32x32 D-UNet to yield the reconstructed HRSI images (Reconstructed). The difference maps are produced by subtracting the reconstructed and ground truth images.

Figure \ref{fig:spec}. Magnitude spectral reconstructions using the three D-UNets are shown for two voxels. The voxel locations for the white matter and lesion spectra are displayed in the T1w image as red points. The spectra generated from the ground truth (red), the 16x16 (blue), 24x24 (green), and 32x32 D-UNets (black) for the 1.5-4.3ppm range are displayed. For spatial comparison, the choline metabolite maps for each method are also shown. All metabolite maps are scaled from 0 to 1. The 24x24 and 32x32 D-UNet reconstructions over-estimate the amount of choline in the lesion. However, the 16x16 reconstruction under-estimates the amount of metabolite signal in the healthy white matter region.

Figure \ref{fig:invivo}. An \textit{in vivo} example of a healthy volunteer is used to demonstrate the potential application for the D-UNet. The experimental high resolution SI (HRSI) data was acquired at 128x128 resolution using an accelerated acquisition protocol\cite{hangel2016ultra}. This data was then down-sampled to produce 32x32 low resolution SI (LRSI) metabolite maps for Cr, NAA, Glu, Gln, Ch, and mI. Together with the T1w image, the low resolution metabolite images were used to reconstruct high resolution spectroscopic images (Recon) using the 32x32 D-UNet model. The difference maps between the Recon and HRSI images (Diff) are also shown. 

Figure \ref{fig:MSE}. The Glu/Cr (A) and Ch/Cr (B) signals averaged over ten subjects are shown for the experimentally acquired images (HRSI) and the images reconstructed from low resolution 32x32 images (Recon). The signals are shown as a function of the T1w intensity, which is representative of the gray and white matter content of the voxel. The red dotted line represents the point at which the gray matter content equals the white matter content in a voxel. The correlation between the HRSI and Recon values are plotted (C) with linear fits. For both Glu/Cr and Ch/Cr, the r$^2$ values of the fits are above 0.99. Finally, the mean squared error for the ten subjects calculated between the HRSI and Recon for each metabolite map (D) is displayed. The dotted black lines reflect the MSE from the testing set for different noise values (low, medium, and high).

\end{document}